\documentstyle[12pt,aasms4]{article}

\def\simless{\mathbin{\lower 3pt\hbox
     {$\rlap{\raise 5pt\hbox{$\char'074$}}\mathchar"7218$}}}   
\def\simmore{\mathbin{\lower 3pt\hbox
     {$\rlap{\raise 5pt\hbox{$\char'076$}}\mathchar"7218$}}}   

\received{}
\accepted{}
\slugcomment{Submitted to Ap.J, May 1999}

\lefthead{Reig et al.}
\righthead{Timing and spectral properties of Aquila X--1}

\begin{document}

\title{Correlated Timing and Spectral Variations of the Soft X-ray Transient 
Aquila X--1: Evidence for an Atoll classification
}

\author{P.~Reig\altaffilmark{1,2},
	M.~M\'endez\altaffilmark{3,4},
        M.~van~der~Klis\altaffilmark{3},
        E.~C.~Ford\altaffilmark{3},
}

\altaffiltext{1}{Foundation for Research and Technology-Hellas, 711 10 
Heraklion, Crete, Greece}

\altaffiltext{2}{Physics Department, University of Crete, 710 03 Heraklion,
Crete, Greece}

\altaffiltext{3}{Astronomical Institute ``Anton Pannekoek'',
       University of Amsterdam and Center for High-Energy Astrophysics,
       Kruislaan 403, NL-1098 SJ Amsterdam, the Netherlands}

\altaffiltext{4}{Facultad de Ciencias Astron\'omicas y Geof\'{\i}sicas, 
       Universidad Nacional de La Plata, Paseo del Bosque S/N, 
       1900 La Plata, Argentina}

\begin{abstract}

Based on {\em Rossi X-ray Timing Explorer} data, we discuss the
classification of the soft X-ray transient Aquila X--1 in the Z/atoll
scheme, and the relation of its kilohertz quasi-periodic oscillations (kHz
QPO) properties to the X-ray colors.  The color-color diagram shows one
elongated ("banana") structure and several "islands" of data points. The
power spectra of the island are best represented by a broken power-law,
whereas those of the banana by a power-law below $\sim$ 1 Hz plus an
exponentially cut-off component at intermediate frequencies (30--60 Hz). 
The parameters of these two components change in correlation with the
position of the source in the color-color diagram. Based on the pattern
that the source shows in the color-color diagram and its aperiodic
variability we conclude that Aquila X--1 is an atoll source.   We have
also investigated the possible correlation between the frequency of the
kHz QPO and the position of the source in the color-color diagram.  The
complexity seen in the frequency {\it versus} count rate diagram is
reduced to a single track when the frequency is plotted against hard or
soft color.

\end{abstract}


\section{Introduction}

The X-ray burst source and soft X-ray transient Aquila X--1 displays X-ray
variability on all time scales.  Usually the source is in its quiescent
state with a very low X-ray luminosity, typically $\sim 10^{33}$  erg
s$^{-1}$ (Verbunt et al. 1994).  Within intervals of months to years it
shows outbursts characterized by a gradual increase in flux to a level
that sometimes is comparable to the Crab, but at other times is two orders
of magnitude fainter. The periodicity of these long-term outburts is
unstable with recurrent periods of $\sim$ 125 and 309 days  (Priedhorsky
\& Terrell 1984; Kitamoto et al. 1993).  The rise time is typically 10
days, the source remains at a flux maximum for 5--10 days and then decays
back to quiescence in typically 30--50 days (Campana et al. 1998; Kitamoto
et al. 1993).  The spectrum is soft and no regular pulsations have been
detected (Cui et al.  1998; Zhang, Yu \& Zhang 1998).  During outbursts,
Aquila X--1 displays Type-I X-ray bursts (Zhang et al.  1998) that can be
explained in terms of runaway thermonuclear burning of matter on the
surface of the neutron star.  As in many other low-mass X-ray binaries
(LMXB), kHz quasi-periodic oscillations (QPO) have been detected in the
persistent flux of Aquila X--1 (Zhang et al.  1998; Cui et al.  1998). 
These kHz QPO occurred in the frequency range 740--830 Hz, had a Q value
(=$\nu_{QPO}/\Delta \nu_{QPO}$) of over 100 and an average fractional
$rms$ amplitude of 7\,\%.  kHz QPO may be caused by orbital motion of gas
around the neutron star very close to its surface (see van der Klis 1999
for a recent review).

Recent deep optical and infrared images and low resolution 
spectroscopy of this system suggest that the mass-donating companion is a
V=21.6 K7V star, located at an estimated distance of 2.5 kpc (Callanan et
al. 1999; Chevalier et al. 1999), and not the V=19.2 K1IV star, 
0.48$^{''}$ West, previously assumed to be the candidate (Chevalier \&
Ilovaisky 1991; Shahbaz, Casares \& Charles 1997).  At radio  wavelengths
Aquila X-1 is unusual in being one of the only neutron star  binaries to
exhibit radio emission (Hjellming \& Han  1995).

The late-type optical counterpart and the episodic outbursts define Aquila
X--1 as a Soft X-ray Transient (SXT), whereas the existence of Type I
bursts denotes a low magnetic field neutron star companion (as opposed to
black hole).  Prior to EXOSAT, low mass X-ray binaries (LMXB) were
classified on the basis of their X-ray luminosity (Schulz, Hasinger \&
Tr\"umper 1989) into two classes:  high and low luminosity systems.  The
low luminosity systems included the so-called X-ray bursters.  Thanks to
its wide orbit and large collecting area, the EXOSAT satellite allowed
long uninterrupted observations and detailed X-ray timing analysis. Taking
into account the rapid aperiodic variability it turned out that low-mass
X-ray binaries could be divided into two different subclasses, Z and atoll
sources, defined by the patterns that these sources display in X-ray
color-color diagrams and the properties of the rapid X-ray variability 
(Hasinger \& van der Klis 1989, hereafter HK89).  The classification of
Aquila X--1 in this scheme is not certain. {\em RXTE} data has suggested
that Aquila X--1 is an atoll source (Cui et al.  1998), though some
properties appear anomalous.  In this paper we investigate the correlated
X-ray timing and spectral variations of  Aquila X--1 and present evidence
for its  classification as an atoll source.

\section{Observations}

The data used in this work were retrieved from the public {\em RXTE}
archive and correspond to two different sets of observations.  The first
one took place between 1997 February 16 -- March 10 and consists of 12
separate observations (one every two days, roughly), with a total usable
time of about 96 ks (see Zhang et al.  1998).  These data correspond to
the decay phase of an outburst.  The observations during the second set,
1997 August 11 -- September 10, were conducted at a typical rate of once
or twice per day and produced approximately 168 ks of usable data.  These
observations began halfway through the rising phase of another outburst
and finished half way through the decay phase.  Five Type I X-ray bursts
(two of them were already reported in Zhang et al. 1998) as well as kHz
QPO were present in the data.  The bursts were excluded from our analysis.
Another four snapshots taken during the rising phase of the 1998
February-March outburst (Cui et al. 1998) were also analysed in order to
compare our results with those of other authors.

{\em RXTE} carries two
pointed instruments, the Proportional Counter Array (PCA) developed to
cover the lower part of the energy range (2--60 keV), and the High Energy
X-ray Timing Experiment (HEXTE) sensitive to X-rays between 15 and 250
keV. These instruments are equipped with collimators yielding a FWHM of
one degree. In addition, RXTE carries an All-Sky Monitor (ASM) that scans
about 80\% of the sky every orbit. In this work we analyzed data from the
PCA only since  its large collecting area ($\sim$ 6500 cm$^2$), makes it
the most appropiate instrument for timing studies.

\section{Analysis and Results}

\subsection{Color-color diagram}

Background subtracted light curves corresponding to the energy ranges
2.0--3.5 keV, 3.5--6.0 keV, 6.0--9.7 keV and 9.7--16.0 keV were used to
define the soft and hard colors as SC = 3.5--6.0/2.0--3.5 and HC =
9.7--16.0/6.0--9.7, respectively.  In a few of the observations one or
two of the five detectors of the PCA were switched off; we only used the
three detectors which were always on to calculate these count rates, and
normalized the count rates to 5 detectors.  The color-color diagram of
Aquila X--1 is shown in Figure~\ref{cd}.

The data points in this color-color diagram fall into several distinct
groups, the hard color being the defining quantity.  The statistical
scatter in the uppermost (open squeres) and middle (open circles) groups
in Figure 1 is considerable.  These groups of points come from
observations at the very end (after March 5th) of the decay of the 1997
March outburst.  The lower branch corresponds to observations from the
rest of the 1997 February--March run and from the entire 1997
August--September run. The points represented with filled squares
correspond to the 1998 March 2 observation. The mean PCA intensity of each
group is given in Table~\ref{fitres}. At the rising part of the 1997
August-September outburst the source moved smoothly from left to right
along the lower elongated branch, and moved back from right to left during
the decay phase. During the 1997 February-March outburst the source also
moved towards lower soft colors as the intensity decreased.

In order to measure the source luminosity in the different states
spanned in the color-color diagram we extracted energy spectra and fitted
an absorbed blackbody plus power-law model to the data. An iron line at
6.4 keV was added if necessary. The 2--10 keV X-ray luminosity, assuming a
distance of 2.5 kpc,  varies from 2.0 $\times$ 10$^{34}$ erg s$^{-1}$ in
the uppermost group to 3.2 $\times$ 10$^{35}$ erg s$^{-1}$ in the middle
one.  The elongated lower group of points presents the highest count
rate.  Here the source luminosity increases from left to right, from 2.4
$\times$ 10$^{36}$ erg s$^{-1}$ to 3.9 $\times$ 10$^{36}$ erg s$^{-1}$.

These structures in the color-color diagram and associated count rate
differences are similar to those of typical atoll sources.  The upper and
middle group of points can be associated with the
so-called island state, whereas the more elongated lower part would then
represent the banana state. Only by looking at the position occupied by
the points of the 1998 observation it is not posssible to tell whether
they define an island or a banana state and an analysis in terms of the
aperiodic variability is needed. In any case they seem to indicate that
the transition between the two spectral states is smooth. The gap between
the branches is probably observational:  no data were obtained for two
days between the banana and the island and for three days between the
island and the extreme island states during the 1997 observations. 

In order to investigate the variability of Aquila X--1 as a function of
the position in the color-color diagram we divided the color-color plane
into several regions as shown in Figure~\ref{cd}.  The two island branches
define the first two groups.  The banana branch was divided so that each
region contains approximately the same amount of data.  We then
approximated the shape of the banana branch with a spline, and used the
parameter $S_{\rm a}$ to measure positions along this spline (M\'endez
et al.  1999).  We set $S_{\rm a}$ to 2 at (SC,HC)= (1.732,0.338) and to
3 at (SC,HC)=(2.045,0.320).  The intermediate positions are obtained
through spline interpolation between the two defining values.  This
could be done for the banana branch only since the discontinuity between
the island and banana branches does not allow to define a unique path.
We arbitrarily assigned the values 1.1 and 1.5 to the two island states.
In this way each one of the regions in which we had divided the
color-color diagram is then characterized by a value of the parameter
$S_{\rm a}$.

\subsection{Noise components}

In order to study the source variability at low frequencies ($\leq$ 512
Hz) we divided the 2--60 keV PCA light curve (no energy selection was
done) of each observation into 256 s segments and calculated the Fourier
power spectrum of each segment up to a Nyquist frequency of 1024 Hz. The
high frequency end (950--1024 Hz) of the power spectra was used to
determine the underlying Poisson noise, which was subtracted before
performing the spectral fitting.  The power spectra were normalized to
fractional $rms$ squared per Hertz (van der Klis 1995). In a few
observations one or two of the five PCA units were switch off.  We have
used only those observations for which the five detector units were
switched on. This implied a loss of $\sim$ 11\% of the data.  To avoid
contributions to the power from the kHz QPOs we restricted the spectral
fitting to the frequency interval 1/256--512 Hz. The 256-s power spectra
were then grouped according to the position of the source in the
color-color diagram: for each region in the banana and each island state
one mean power spectrum was obtained.

The power spectra were fitted using two broad noise components 
called the very-low frequency noise (VLFN) and the high-frequency noise
(HFN).  These two components are mathematically represented by a power-law
and a power-law times an exponential cut-off, respectively (HK89).  The VLFN
accounts for the low-frequency part of the spectrum, whereas the HFN
dominates at higher frequencies.  Figure~\ref{pds} shows six power spectra
corresponding to different positions in the color-color diagram; all the 
island states and the lower, middle and upper banana states are shown.

While the VLFN plus HFN model can satisfactorily describe the
banana-state power spectra, it does not provide good fits for the
faintest ($S_{\rm a}=1.1$) of the two island-state power spectra.  Ford
\& van der Klis (1998) used a broken power-law plus one low-frequency
(10--50 Hz) Lorentzian, possibly representing Lense-Thirring precession
(Stella \& Vietri 1998), to fit the island-state power spectra of the
atoll source 4U 1728--34.  In Aquila X--1 we find that a simple broken
power-law gives good fits ($\chi^2_{\nu}$ $\leq$ 1.2) to the power
spectra in both island states.  Although described by different mathematical 
functions, the cut-off power-law of the banana state and the broken power-law of 
the island state are likely to represent the same type of noise component, 
namely HFN. To be consistent, we have used the broken power-law model
to fit the power spectra of the island states.  Figure~\ref{rms}
shows the variation of the amplitude (as fractional $rms$) of the VLFN
(circles), HFN (squares) and broken power-law (triangles) components.
The latter refers to the island states.

Unlike the HFN component, the amplitude of the VLFN component does not
show any clear trend, but remains at $\sim$ 6-7\% $rms$, irrespective of
the position that the source occupies in the banana. The maximum strength
of the HFN is found during the island phases ($rms=$ 34\% and 18\% for the
upper and lower island sates, respectively).  This is in contrast to the
results of Cui et al.  (1998) who did not detect such HFN component in the
first observations of the 1998 February-March outburst (MJD 50875--77),
despite the inference from the color diagrams that these data were in an
island state. These authors discuss the absence of an HFN in the island
state as an unusual phenomenon. We have reanalysed the first four
observations in the Cui et al. (1998) paper and found that there is, in
fact, an HFN component with an $rms$ of 6.0$\pm$0.3\% (Fig~\ref{pds},
Table~\ref{fitres}) in the first two observations.   As expected in the
island, there is no VLFN (1.6\% upper limit, 95\% confidence level).  The
mean hard and soft colors of the corresponding points are HC=1.84, SC=0.34
and HC=1.86, SC=0.31 for the first and second observation, respectively.
That is, in our color-color diagram they lie in between the lower island
and the banana states. This distribution of the island points in the
color-color diagram suggests a continuous transition between the island
and banana groups. The other two observations produced points in the
banana only.

\subsection{kHz QPOs and the color-color diagram}

For the kHz QPO analysis we produced power spectra using 64 s data
segments and a Nyquist frequency of 2048 Hz.  kHz QPO were only observed
in a specific range in the color-color diagram, namely the lower banana
near $S_{\rm a} = 2.32$ (the filled circles in Fig.~\ref{cd}).  The
2--60 keV fractional amplitudes of the kHz QPO ranged from $\sim$ 4.5\%
to $\sim$ 11.7\%.  We did not detect QPO in the upper part of the
banana, with a 95\% confidence upper limit of 1.4\% $rms$ nor in the
island with an upper limit of $\sim$ 8\%.

Figure~\ref{figrate} shows the dependence of the frequency of the QPO as
a function of the 2--16 keV PCA count rate.  The two parallel groups at
the lower left of the plot (open and grey circles) correspond to the
1997 February-March observations (Zhang et al.  1998a), while the rest
come from the 1997 August-September set of observations.

M\'endez et al.  (1999) found that there is a much better correlation
between the frequency of the kHz QPO in 4U 1608--52 and the position of
the source on the color-color diagram than between frequency and flux. 
Similarly, the multi-valued dependence of $\nu_{\rm QPO}$ on X-ray flux
in Aquila X--1 is reduced to a single relation when $\nu_{\rm QPO}$ is
plotted against the $S_{\rm a}$ parameter, which in our case is just a
measure of the soft color (Fig.~\ref{fig_hc}).  The hard color is not as
sensitive to changes in the kHz QPO frequency in Aquila X--1 as it is in
4U 1608--52 (Kaaret et al.  1998; M\'endez et al.  1999).

\section{Discussion}

We have measured the color and timing properties of the  LMXB Aquila X--1,
and found a behavior similar to that of other low-luminosity LMXBs.  The
color-color diagram shows the classical atoll shape with the banana and
island states. The island state is not defined by a single group of points
but it is split into a number of different groups as in other atoll
sources, e.g. 4U 1636--53 (Prins \& van der Klis 1997) and 4U 1608--52
(Yoshida et al. 1993). The lowest count rate and fluxes are detected in
the island state with the hardest color and increase as the hard color
decreases. 

Based on the similarity of the color-color diagram of Aquila X--1 to that
of other atoll sources it seems likely that Aquila X--1 can be placed in
the group of atoll sources.  However, the information provided by the
color-color diagram is, by itself, not always sufficient to classify the
source state.  It is not always possible to make a distinction between Z
and atoll sources or between the two spectral states (island/banana)
within the atoll class (HK89).  Moreover, in some atoll sources the island
and banana branches have been seen to shift in the color-color diagram,
both in soft and hard colors (Prins \& van der Klis 1997).  The
information obtained from the analysis of the noise components in the
power spectra provides the key for an unambiguous classification.

The preliminary classification of Aquila X--1 as an atoll source on the
basis of the color-color diagram is confirmed by the fast-timing
analysis.  The power spectra of atoll sources (HK89) are characterized by
two broad noise components called the very-low frequency noise  and the
high-frequency noise.  The relative strengths of these two components vary
in anticorrelation with each other and with the inferred mass accretion rate,
$\dot{M}$, as measured by $S_{\rm a}$ in our analysis:  the VLFN component
appears at the highest inferred $\dot{M}$, whereas that of the HFN
decreases as $\dot{M}$ increases. These two components are present in the
power spectra of Aquila X--1 (see Fig.~\ref{pds} and Table \ref{fitres}). 
As is commonly seen in atoll sources the VLFN is most prominent in the
banana state, while the HFN dominates the island state.  The fractional
amplitude of the HFN component decreases as the system moves from the
island state to the lower banana and from here to the upper banana.  The
VLFN component is practically undetectable ($rms$ $\leq$ 2\%) in the
island state, and although present in the banana state its amplitude
does not change as expected (see below).  Another similarity between
the power spectra of Aquila X--1 in the banana state and those of atoll
sources is the presence of wiggles in the VLFN (HK89).

The increase of the break frequency as the X-ray count rate (or flux)
increases, and the fact that the largest fractional amplitude and the
hardest spectrum are observed in the two island states, represent
further evidence in favor of an atoll classification for Aquila X--1
(e.g.  4U 0614+09, M\'endez et al.  1997).  It is worth noting that all
these characteristics are also seen in black hole candidates during the
low state, emphasizing the observational similarities between atoll
sources and black holes systems (van der Klis 1994).

One of the open questions in atoll sources is whether the transition from
the island to the banana states occurs continuously or abruptly, that is, 
with the source jumping from one state to the other. The trends of the
spectral parameters just described seem to indicate that such transition
is continuous. This idea would be supported by the 1998 March data (Cui et
al. 1998).  The upper limit to the $rms$ of the VLFN component in the 1998
March observation ($< 1.6$\%, 95\% confidence) is lower than the value in
the banana (5--7\%), and the $rms$ of the HFN is larger (6\% compared to
3.5\%).  The break frequency $\nu_{br}$ $\approx$ 43 Hz follows the same
trend as the other two island states, namely, increases as the source
moves into the banana.

An interesting difference between the island state of 1998 March and the
other two island states is the source count rate. While the  two 1997
March islands show the lowest PCA intensity, the 1998 March one is
comparable to that detected in the lower banana state. It is worth
noting that the 1997 island data were collected during the {\em decay} of an
X-ray outburst, whereas that of 1998 correspond to the {\em rise} of the
outburst. A hysteresis effect may be present.

Our observed VLFN amplitude does not change significantly in the banana
state, contrary to what is observed in other atoll sources.  Also, the
power-law index of the VLFN becomes less steep as the source moves up in
the banana branch, while usually in other atoll sources a slight increase,
if any change, occurs.  This unusual behavior of the VLFN component may be
related to nuclear burning on the surface of the neutron star.  In
periodic X-ray burters, like Aquila X--1, the entire surface of the star
is rapidly ($\leq$ 10 s) burned by a fast propagating thermonuclear
instability. This can only happen at low mass accretion rates, where the
envelope is convectively combustible most of the time. The convective
burning makes it difficult for slower combustion to occur, thus
suppressing the VLFN (Bildsten 1993). Yu et al.  (1999) reported another
unusual aspect of the VLFN in Aquila X--1, namely, its disappearance after
a Type I burst, in which the 2--10 keV flux decreased by about 10\% and
the kHz QPO frequency fell abruptly by $\sim$ 37 Hz.  

Aquila X--1 is unusual among the group of low-mass X-ray binaries with kHz
QPO in that only one kHz QPO has been so far observed.  All other sources
(Z and atoll) have at least sometimes shown two simultaneous kHz QPOs. 
Nevertheless, the relationship between the frequency of the kHz QPO and
the X-ray flux in Aquila X-1 is very similar to that seen in the atoll
sources 4U 1608--52 (M\'endez et al.  1999) and 4U 0614+091 (Ford et al. 
1997):  on time scales longer than $\sim$ 1 day there is no correlation
between the kHz frequency and the X-ray intensity (or flux), whereas on
short time scales ($\sim$ hours), these two quantities correlate
remarkably well (Zhang et al.  1998). We also find that similarly to
other sources, on time scales longer than a day, the QPO frequency
correlates much better to color (or $S_{\rm a}$) than to count rate.

\section{Conclusion}

We have used {\em RXTE} data to carry out a timing analysis with the aim
of solving the issue of the classification of Aquila X--1 in the Z/atoll
scheme.  Aquila X--1 traces a pattern in the color-color diagram that is
consistent with having an island and a banana branch.   The PCA intensity
correlates with the position of the source in the color--color diagram,
decreasing in the sense upper banana $\longrightarrow$ lower banana
$\longrightarrow$ island state.

The power density spectra of the banana states can be described in terms
of a power-law component, the so-called very low-frequency noise and a
cut-off power-law component or high-frequency noise.  In the island state
the high frequency noise component is best described in terms of a broken
power-law, whereas the very low frequency noise is practically
undetectable. The characteristics of the power spectra also change in
correlation with the position of the source in the color-color diagram.
The amplitude of the high frequency noise increases as the $S_{\rm a}$
parameter decreases. Finally, we have shown that the soft color (via the
$S_{\rm a}$ parameter) correlates with the frequency of the kHz QPO.

Based on the correlation of the X-ray spectral properties of Aquila X--1
(its position in the color-color diagram) and its fast-timing behavior we
conclude that Aquila X--1 is an atoll source.

\acknowledgements

This work was supported by the Netherlands Foundation for research in
astronomy (ASTRON) under grant 781-76-017, by the Netherlands Research
School for Astronomy (NOVA), and the NWO Spinoza grant 08-0 to E.P.J.
van den Heuvel.  PR acknowledges support from the European Union through
the Training and Mobility Research Network Grant ERBFMRX/CT98/0195.  MM
is a fellow of the Consejo Nacional de Investigaciones Cient\'{\i}ficas
y T\'ecnicas de la Rep\'ublica Argentina.  This research has made use of
data obtained through the High Energy Astrophysics Science Archive
Research Center Online Service, provided by the NASA/Goddard Space
Flight Center.

\clearpage

\begin{deluxetable}{cccccccccc}
\scriptsize
\tablecolumns{10}
\tablecaption{Power spectral parameters \label{fitres}}
\tablewidth{0pt}
\tablehead{}
\startdata
\multicolumn{9}{c}{Island State} \\
& \multicolumn{2}{c}{VLFN} &  & \multicolumn{4}{c}{HFN} & & \\
\cline{2-3} \cline{5-8} \\
$ S_{\rm a}^a$ & rms$^g$ & $\alpha$  & & rms$^b$ & $\alpha_1^c$ & $\nu_{\rm br}^d$ & $\alpha_2^e$ 
& count rate$^f$ & $\chi^2_{\nu}$(dof) \\
\tableline
1.1 & \nodata & \nodata & & $34 \pm 2$ & $0.18^{0.16}_{0.25}$ & $0.14 \pm 0.05$ 
& $0.63 \pm 0.04$ & 45	&1.23(82) \\
1.5 & \nodata & \nodata & & $18.4 \pm 0.5$ & $-0.01 \pm 0.06$ & 
$15.7^{+0.6}_{-1.6}$ & $1.22 \pm 0.11$ & 200 & 0.79(82) \\
1998 March 2& $< 1.6$ &1.5 (fixed) && $6.0 \pm 0.3$ & $0.06 \pm 0.10$ &
$43^{+12}_{-4}$ &
$1.9 \pm 0.6$ & 1481  & 0.80(112)  \\
\tableline\\
\multicolumn{9}{c}{Banana State} \\
& \multicolumn{2}{c}{VLFN} &  & \multicolumn{4}{c}{HFN} & & \\
\cline{2-3} \cline{5-8} \\
$ S_{\rm a}^a$ & rms$^g$ & $\alpha$ & & rms$^b$ & $\alpha$ & $\nu_{\rm cut}$ & & 
count rate$^f$ & $\chi^2_{\nu}$(dof) \\
\tableline
$2.32 \pm 0.10$ & $5.6 \pm 0.2$ & $1.61 \pm 0.03$ & & $3.5 \pm 0.1$ & $-0.7 \pm 
0.2$	
& $35^{+9}_{-7}$ & \nodata &1166	& 1.67(81) \\
$2.60 \pm 0.08$ & $7.2 \pm 0.2$ & $1.59 \pm 0.02$ & & $3.0 \pm 0.1$ & $-0.6 \pm 
0.2$	
& $31^{+8}_{-6}$ & \nodata &1629	& 1.99(81)  \\
$2.72 \pm 0.06$ & $6.2 \pm 0.2$ & $1.55 \pm 0.03$ & & $2.9 \pm 0.1$ & $-0.2 \pm 
0.2$	
& $64^{+21}_{-17}$ & \nodata &1977	& 1.11(81) \\
$2.82 \pm 0.06$ & $5.7 \pm 0.2$ & $1.46 \pm 0.2$  & & $2.1 \pm 0.2$ & 
$-0.7^{+0.4}_{-0.7}$	
& $27^{+12}_{-8}$ & \nodata &2013	& 1.02(81) \\
$2.92 \pm 0.06$ & $6.3 \pm 0.1$ & $1.49 \pm 0.2$  & & $2.2 \pm 0.1$ & $-0.4 \pm 
0.2$	
& $34^{+11}_{-8}$ & \nodata &2169	& 1.46(81) \\
$3.04 \pm 0.08$ & $5.6 \pm 0.1$ & $1.41 \pm 0.02$ & & $1.8 \pm 0.1$ & 
$-0.5^{+0.2}_{-0.3}$	
& $31^{+17}_{-10}$ & \nodata &2342	& 1.41(81)  \\
\tableline
\enddata
\tablenotetext{}{In the island state the HFN is represented by
a broken power law; in the banana state the HFN is represented
by a power law with an exponential cut-off (see text).
Errors are based on a scan in $\chi^2$ space using $\Delta \chi^2=1$.}
\tablenotetext{a}{Refers to the regions defined in the color-color diagram
(see text)}
\tablenotetext{b}{1 -- 100 Hz rms amplitude (\%).}
\tablenotetext{c}{Broken power law; slope below the break.}
\tablenotetext{d}{Broken power law; break frequency (Hz).}
\tablenotetext{e}{Broken power law; slope above the break.}
\tablenotetext{f}{Background subtracted count rate for the full PCA (c/s).}
\tablenotetext{g}{0.001 - 1 Hz rms amplitude (\%).}
\end{deluxetable}

\clearpage

\figcaption[fig1.ps]{
Color-color diagram of Aquila X--1.  The soft and hard colors are
defined as the ratio of count rates in the bands $3.5 - 6.0$ keV and
$2.0 - 3.5$ keV, and $9.7 - 16.0$ keV and $6.0 - 9.7$ keV, respectively.
The contribution of the background has been subtracted, but no dead-time
correction was applied to the data (the dead-time effects on the colors
are less than 1\%).  Each point in the banana branch (closed circles) and
in the island with the lowest hard color (filled squares) represents 64 s of
data, 128 s in the middle island (open circles), and 768
s in the extreme island (open squares).  We show the typical error bars
in the banana and the island states.  Black and grey symbols indicate
segments with and without kHz QPOs, respectively.  Vertical lines define
the regions into which the banana branch was divided.  
\label{cd} }

\figcaption[fig2.ps]{ Power spectra with best fits for six different
positions in the color-color diagram:  $a)$ extreme island, $b)$ island,
$c)$ lower island (1998 March observation) $d)$ lower banana ($S_{\rm
a}=2.32$) $e)$ middle banana ($S_{\rm a}=2.82$) and $f)$ upper banana
($S_{\rm a}=3.04$), see Table \ref{fitres}.  The Poisson level has been
subtracted. \label{pds} }

\figcaption[fig3.ps]{
VLFN (circles) and HFN (squares and triangles) fractional rms (1--100
Hz, full energy band) as a function of the $S_{\rm a}$ parameter (see
text).  The two triangles correspond to the island and extreme island
state and were obtained using a broken power-law function.  The $S_{\rm
a}$ values for these two points were arbitrarily chosen.
\label{rms}
}

\figcaption[fig4.ps]{
Relation between the frequency of the lower kHz QPO and the $2 - 16$ keV
count rate.  The count rates have been corrected for background.  Each
point represents a 128-s segment.  Different symbols represent different
observing times as follows:  MJD 50506.245--50506.358 (grey circles),
MJD 50508.898--50508.993 (open circles), MJD 50673.004--50673.473
(filled circles), MJD 50675.793--50675.809 (open squares), MJD
50677.254--50677.262 (filled triangles), MJD 50697.519--50697.684
(filled squares).
\label{figrate}
}

\figcaption[fig5.ps]{
Relation between the frequency of the lower kHz QPO and $S_{\rm a}$ (see
Fig.  \ref{cd}), for the same segments and symbols as those shown in
Figure \ref{figrate}.
\label{fig_hc}
}

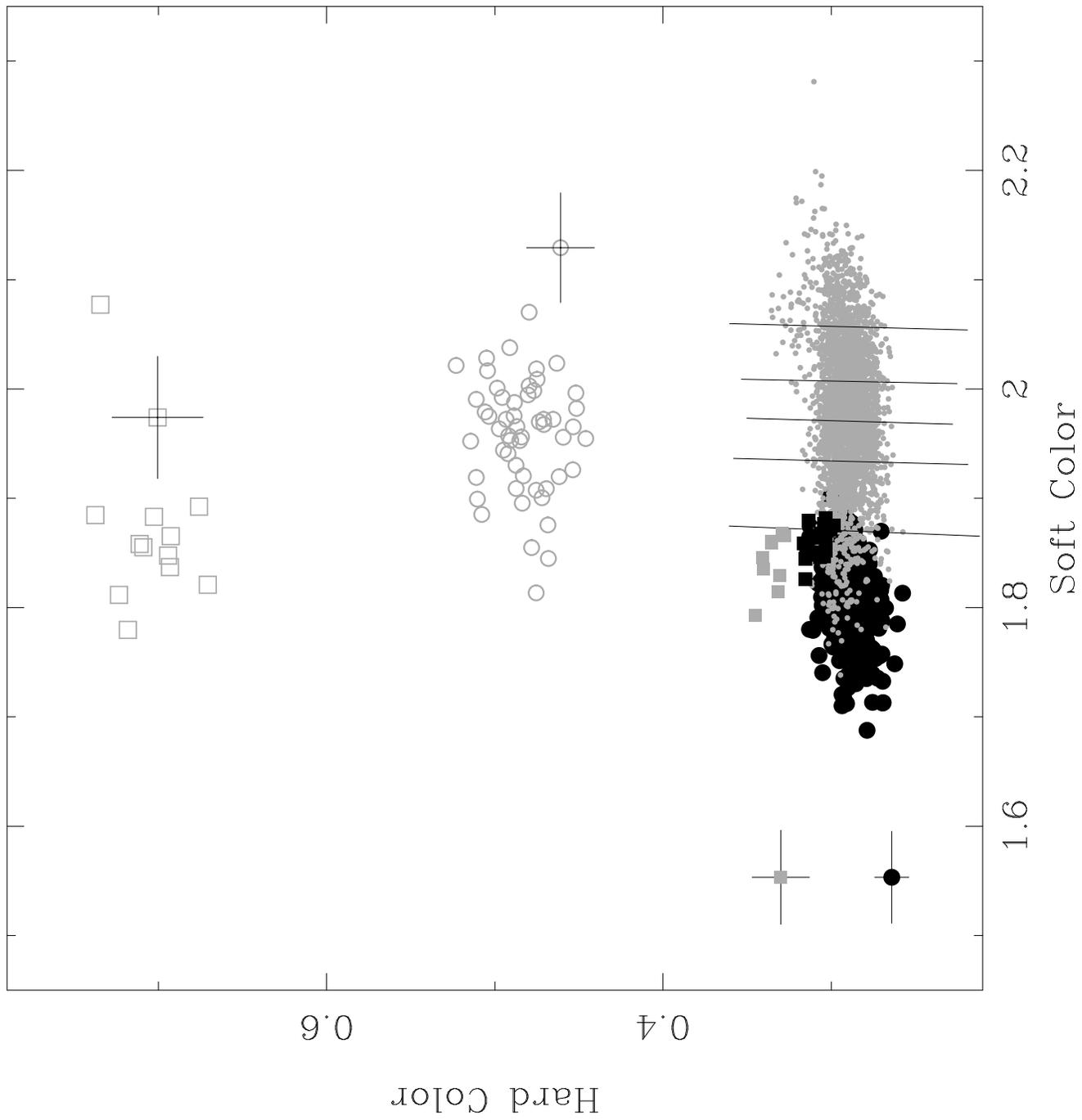
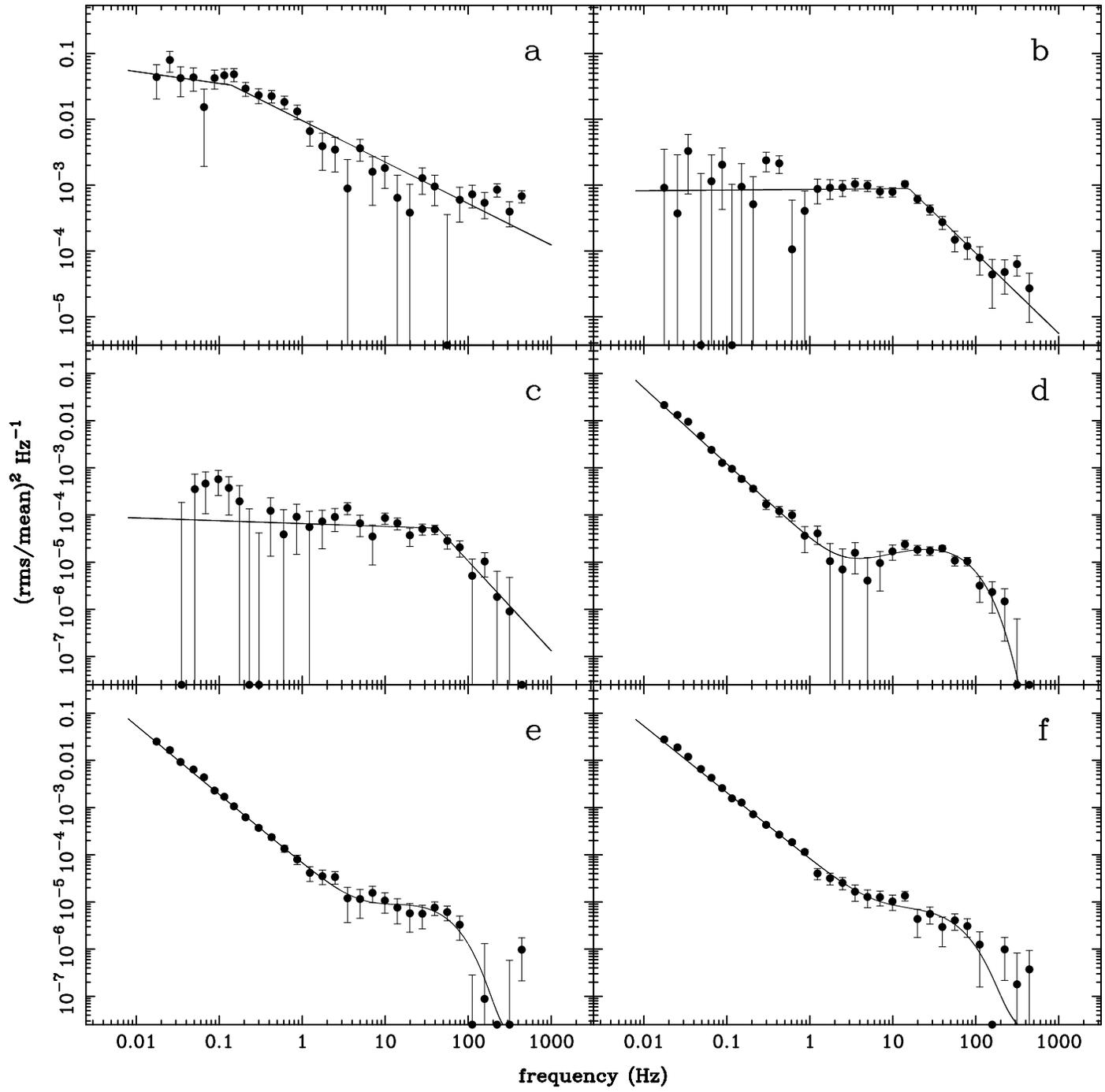
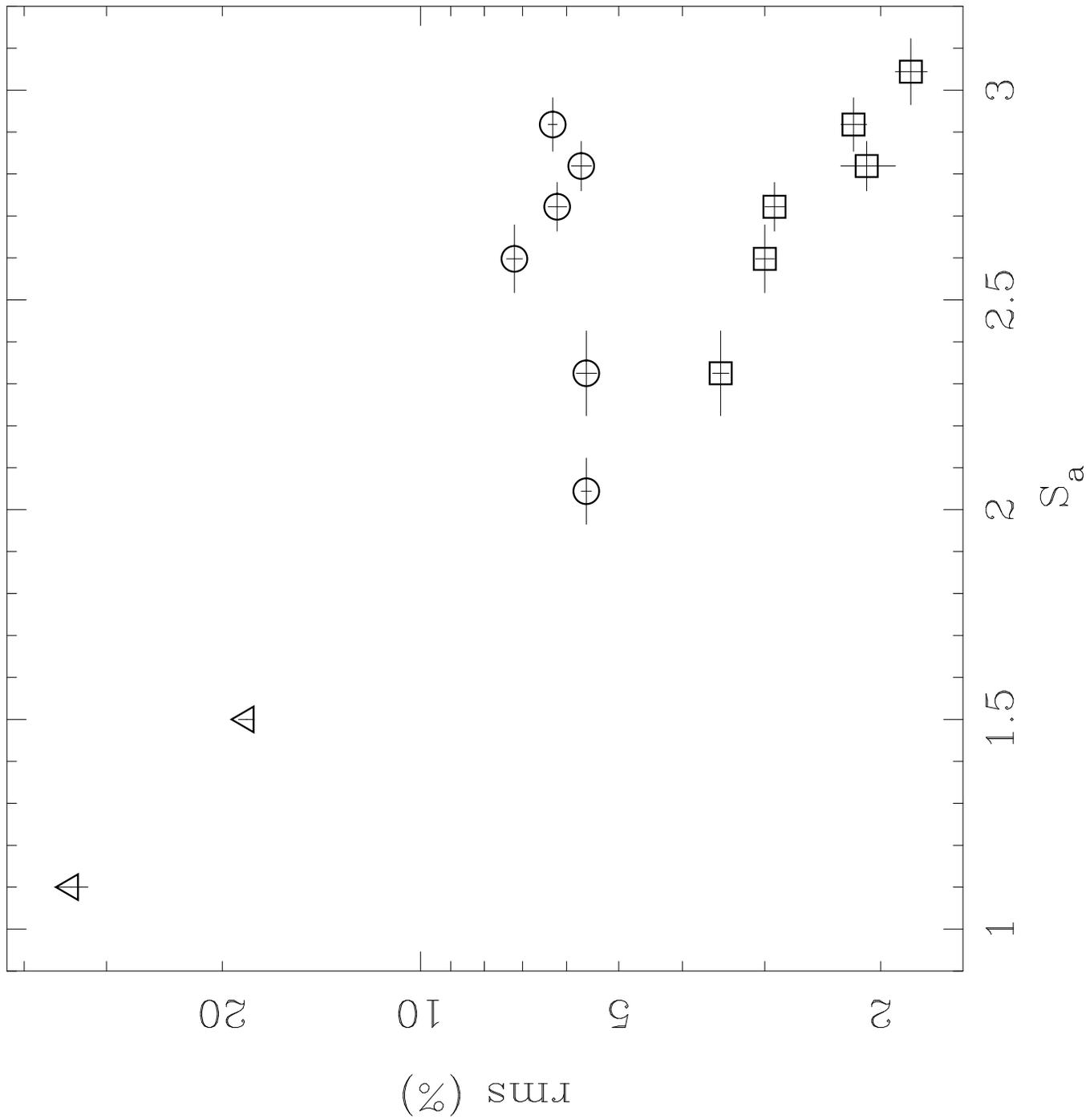
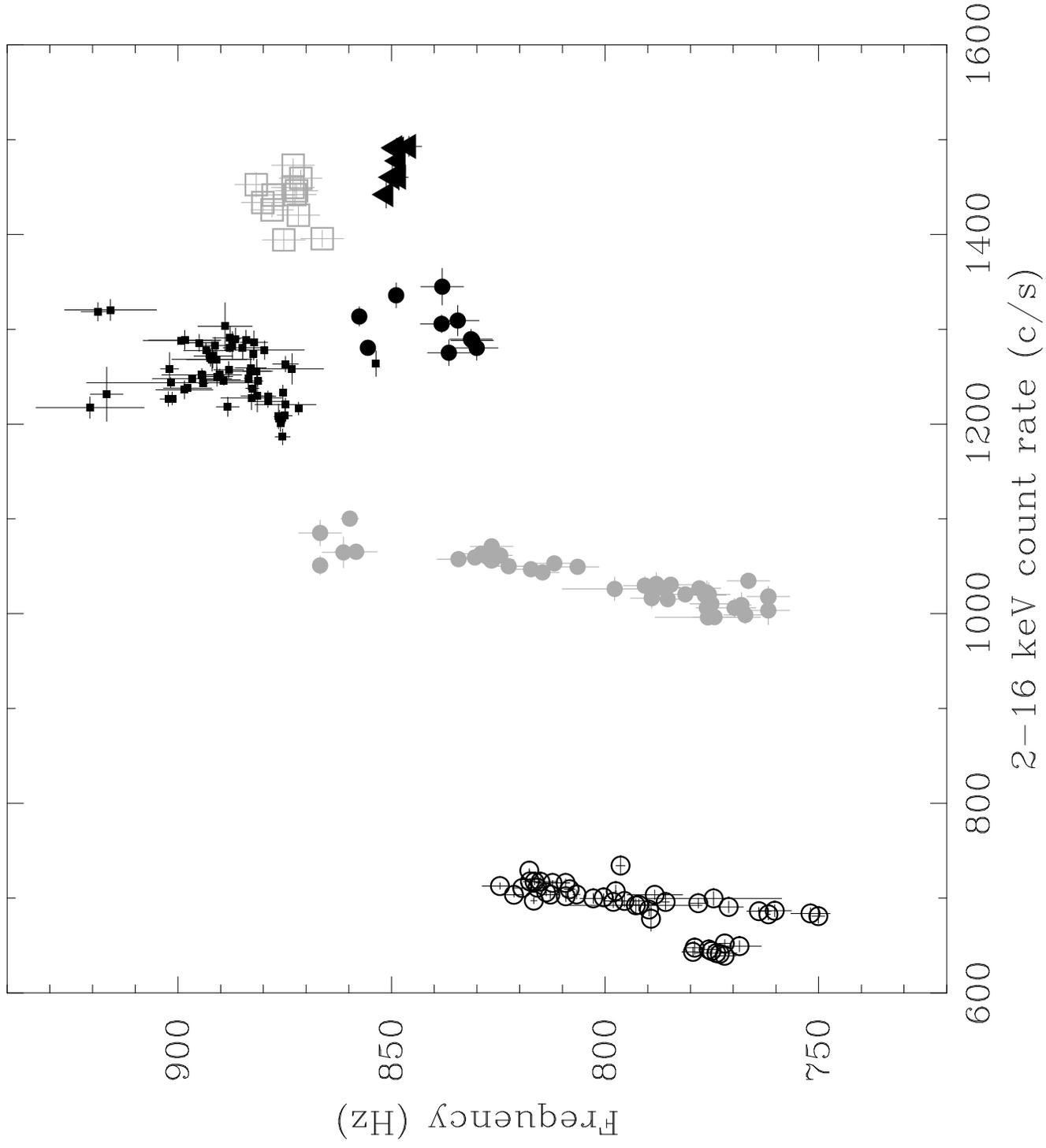
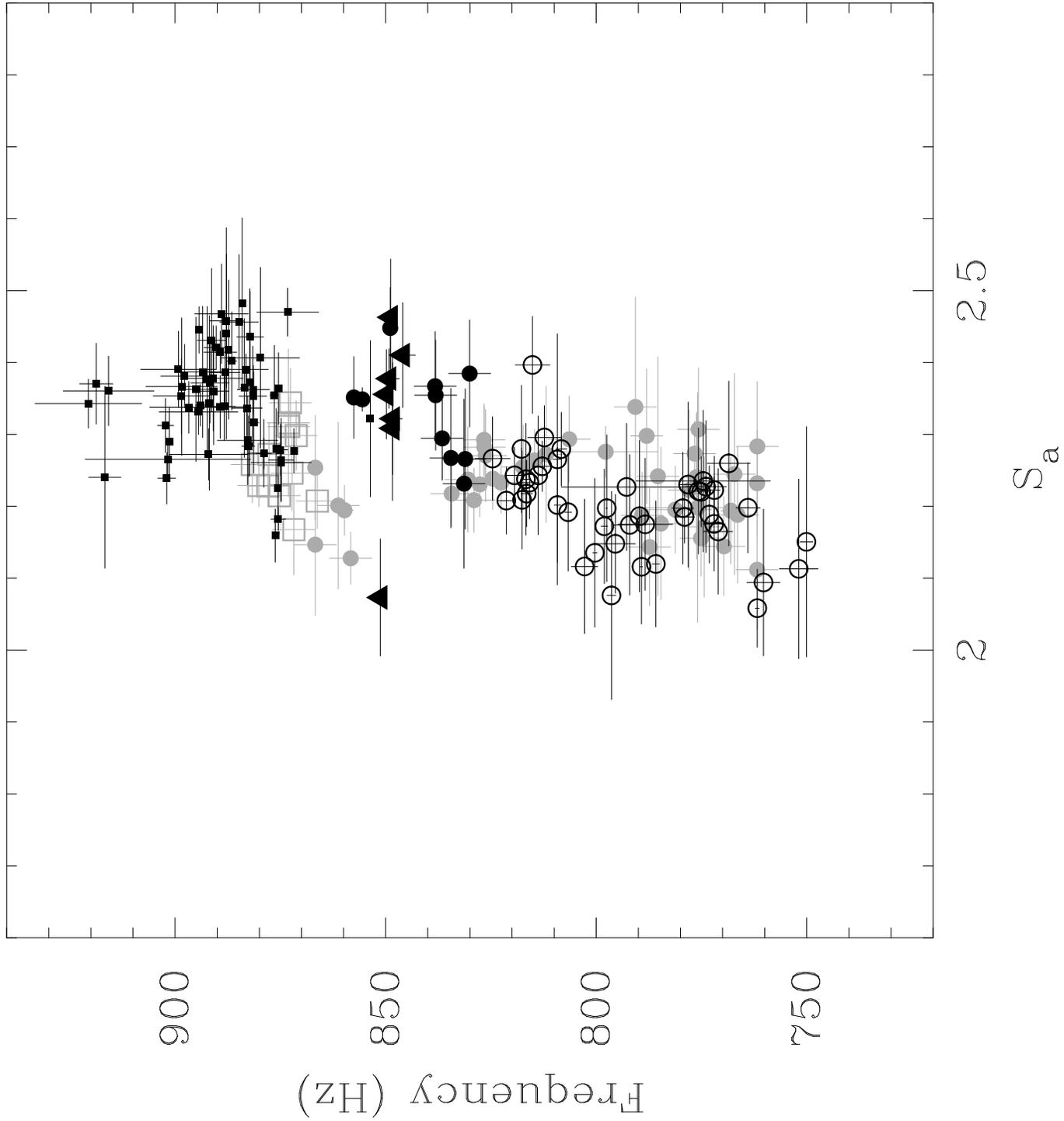
\end{document}